\documentclass[12pt]{article} 
\usepackage{epsf}  
\usepackage{float}  
\begin{document} 
\baselineskip 14pt plus 2pt minus 2pt 
 
\newcommand{\no}{\nonumber}  
\newcommand{\bea}{\begin{eqnarray}}  
\newcommand{\eea}{\end{eqnarray}}  
\newcommand{\lanl}{\langle}  
\newcommand{\ranl}{\rangle}  
\newcommand{\del}{\partial} 
\renewcommand{\thefootnote}{\#\arabic{footnote}}
 
\begin{titlepage} 

\hfill {\tiny FZJ-IKP(TH)-2001-01}

\vspace{1cm}

\begin{center} 
 
{\large  \bf { 
Complete analysis of pion--nucleon scattering \\[0.3em] 
in chiral perturbation theory to third order}}

\vspace{1.2cm} 
                               
{\large Nadia Fettes\footnote{email: N.Fettes@fz-juelich.de}, 
Ulf-G. Mei\ss ner\footnote{email: Ulf-G.Meissner@fz-juelich.de} 
}

\vspace{1.0cm} 
 
{\it Forschungszentrum J\"ulich, Institut f\"ur Kernphysik (Theorie)} 
 
{\it D--52425 J\"ulich, Germany}\\

\end{center}

\vspace{0.8cm}

\begin{abstract} 
\noindent  
We consider pion--nucleon scattering in heavy--baryon chiral  
perturbation theory to third order.  
All electromagnetic corrections appearing to this order are  
included. We thus have a consistent description of strong and  
electromagnetic effects, which allows us to isolate the strong part  
of the interaction in an unambiguous way. We give pion--nucleon phase shifts up to pion 
laboratory momenta of 100~MeV and find sizeable differences for the S--waves of 
the elastic channels compared to previous phase shift analyses. The precise description of the scattering 
process also allows us to address the question of isospin violation 
in the strong interaction. For the usually employed triangle relation 
we find an isospin breaking effect of  $-0.7\%$ in the S--wave, 
whereas the P--waves show an effect of $-1.5 \%$ and  
\hbox{$-4$ to $-2.5 \%$}, respectively, for  pion laboratory momenta
between 25 and 100~MeV. 
\end{abstract}

\vspace{2cm}
\small{

}

\vfill 
 
\end{titlepage}

\section{Introduction 
\label{sec:intro}} 
 
\medskip\noindent 
The study of the strong interactions at low energies has stimulated vast efforts 
in the domain of pion--nucleon physics. It is mainly the description 
of simple processes like pion--nucleon scattering which can help us  
to extend our knowledge about the fundamental principles of the 
strong interactions. For every field theory, the consideration of symmetry  
principles is very important, and it is thus of primordial interest 
to know to what extent these symmetries are realized in nature. 
 
\medskip\noindent 
For a long time, isospin symmetry has been thought of as an exact symmetry of  
the strong interactions. But the nucleon isospin doublet, consisting of proton  
and neutron, is not degenerate in mass; whereas electromagnetic interactions  
alone would make the proton heavier than the neutron, the mass difference between the 
up-- and down--quarks reverses the situation: the neutron turns out to be heavier than the 
proton by 1.3~MeV. This simple consideration triggers questions about the 
size of isospin violation in other strong processes. 
 
\medskip\noindent 
Pion--nucleon scattering is the ideal playground for studying the question of  
isospin breaking. Indeed, thanks to the construction of meson factories and the  
formidable effort of many experimental groups, we now have a large amount of 
very precise low--energy data. Weinberg pointed out on purely theoretical  
grounds that isospin breaking effects in neutral--pion scattering must be  
dramatically enhanced due to the smallness of the isoscalar amplitudes~\cite{wein77}.  
The advent of high--quality data opened the way to the study of isospin symmetry  
violation in processes which are experimentally accessible: these are the  
elastic $\pi^+ p$ and $\pi^- p$ as well as the single charge exchange (SCX) reactions.  
The difficulty now lies in the extraction of the strong interaction amplitude 
from the scattering data. Electromagnetic corrections have to be 
applied in order to gain information about the hadronic amplitudes. 
Such corrections have been computed e.g.\ by the Nordita group~\cite{nordita} 
and more recently by Gashi et al.~\cite{gashi}. The problem   
is that these analyses cannot account for the electromagnetic and the strong interaction  
in a consistent way. In view of studying small quantities like isospin symmetry 
violation, it is absolutely necessary to avoid systematic errors resulting from 
a possible mismatch in the description of both forces. We will readdress this question,  
i.e.\ we will extract the strong amplitude from scattering data in 
the framework of chiral perturbation theory ($\chi$PT). $\chi$PT allows to 
calculate the electromagnetic and the strong interactions consistently in terms 
of unknown parameters, the so--called low--energy constants (LECs). Having 
fitted the full amplitude to the available cross section data, one is then able 
to switch off electromagnetism and to predict the strong piece in a unique fashion. 
 
\medskip\noindent 
The knowledge of the hadronic amplitude is necessary for analyzing the violation 
of isospin symmetry of the strong interaction. Such analyses have been performed  
in the past~\cite{gibbs,matsi}, indicating breaking at low energies as large as $6 \%$. Both  
works, however, are based on electromagnetic corrections which might not be compatible 
with the description of the strong interactions. In order to better understand the origin 
of this large isospin breaking effect, we readdress this question within our framework. 
The advantage of such a  diagrammatic calculation lies in the fact that we can 
clearly localize which mechanisms are responsible for isospin violation, thus making  
a detailed study of the subject feasible.

\section{The amplitude 
\label{sec:amplitude}} 
 
This work follows a series of previous papers: in~\cite{FMS}, pion--nucleon scattering  
was studied in the limit of isospin symmetry. The unknown counterterms were determined 
by a fit to phase shifts resulting from three different analyses~\cite{ka85,mats,sp98}.  
In~\cite{fmsi,fmi}, we included strong isospin breaking terms in the amplitude and took  
the full mass difference of the physical particles into account. The isospin symmetric  
amplitude had to be generalized in order to describe scattering of particles with  
different masses. Not only do the particles in the initial and final states have different masses, 
but pion mass differences also have to be considered in the loops, and the nucleon  
mass difference plays a role in the intermediate states. 
 
\begin{figure}[htb] 
\centerline{ 
\epsfysize=1.8in 
\epsffile{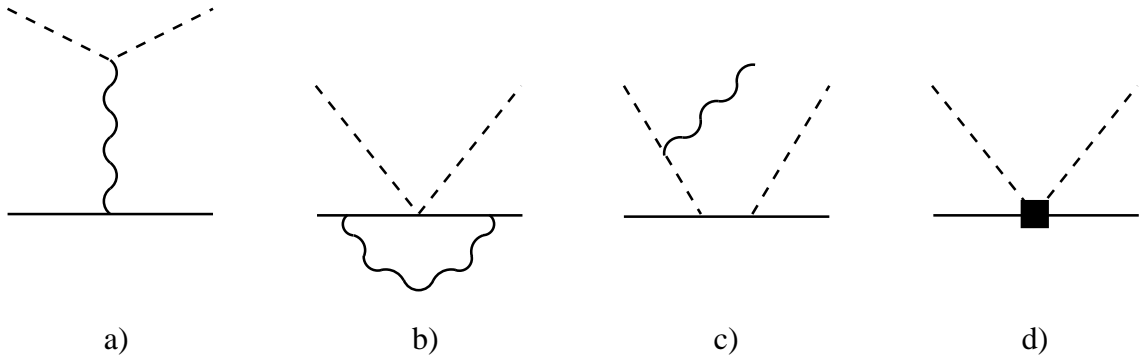} 
} 
\vskip 0.5cm 
\caption{ 
Example diagrams for a)~the Coulomb potential, b)~a ``usual'' photon loop, c)~Bremsstrahlung, 
and d)~an electromagnetic counterterm. The solid, dashed, and wiggly lines denote nucleons,  
pions, and photons, in order.} 
\label{fig:photo} 
\end{figure}

\medskip\noindent 
The aim of this work is to describe QCD and QED with unequal up-- and down--quark  
masses ($m_u \neq m_d$) and non--vanishing electric charge ($e^2 \neq 0$). 
We thus have to add loops with virtual photons as well as electromagnetic  
counterterms. This is the yet missing part in the {\em complete} analysis of pion--nucleon  
scattering. 
Photons enter the amplitude via four mechanisms, one example of each is shown in  
fig.~\ref{fig:photo}. 
 
\begin{itemize} 
\item  
First, there is the Coulomb potential, which is the exchange 
of soft photons between the charged pions and the proton. To the 
accuracy we are working we only have to include the one-- and two--photon exchange, 
since multi--photon exchange diagrams are suppressed by higher powers of $e^2$. 
It is important to note that our simultaneous counting of momenta $q$ and  
electromagnetic couplings $e$ as small  
quantities necessarily breaks down at very low energies. To illustrate this,  
let us compare the Weinberg--Tomozawa term (which in our counting is of first order) 
and the same process accompanied by the exchange of a photon between  
the initial or the final particles. At low energies, the loop  
diagram (which is of third order) is ``suppressed'' by a factor  
$e^2 M_\pi/ (32 |\vec{q}_\pi|)$. Thus, for pions of momentum less than 
$0.5$~MeV, the loop diagram is not suppressed compared to the tree diagram, 
which clearly contradicts our counting scheme. At very low energies, one 
would thus have to sum up diagrams with an arbitrary number of photons 
exchanged between the charged--particle legs. This is commonly done in 
nucleon--nucleon scattering, where the heavy nucleons can be considered 
to be non--relativistic at such energies. Fortunately we do not have to  
worry about these ladder exchange diagrams since they only become  
important at energies much lower than the experimentally accessible energy range.  
In our analysis, we will thus stick to the traditional 
chiral counting, which consists in considering $e$ a quantity of chiral order $q$. 
\item 
Second, there are the ``usual'' loop diagrams with virtual photons, i.e.\ those  
which include 
at least one strong--interaction vertex. Many of these only renormalize the coupling constants. 
\item 
Third, we also have to consider Bremsstrahlung processes where a soft photon is emitted 
from an external particle, but is not detected since its energy is lower than the 
detector resolution $\Delta E$. The inclusion of this process makes  
the amplitude infrared finite. We have not performed this calculation 
analytically, since it is sufficient for the actual computation to  
combine every $\log(m_\gamma/M_\pi)$ from virtual--photon diagrams with  
$\log(\Delta E/m_\gamma)$ due to the integrated emission of soft photons 
with energy smaller than the detector resolution. The divergence appearing because of 
the vanishing photon mass $m_\gamma$ thus cancels out. For the evaluation, 
we choose $\Delta E = 10$~MeV\footnote{This choice is rather arbitrary;  
we have checked, however, that our results do not depend on the particular choice of $\Delta E$}. 
\item 
Fourth, hard photons have been integrated out of the theory. Their effect  
is parameterized by a string of terms in the Lagrangian, each accompanied 
by an unknown coupling constant. These constants, similarly to the 
hadronic counterterms, have to be fitted to experimental data. 
\end{itemize} 
 
\medskip\noindent 
The inclusion of photons and electromagnetic counterterms in the  
amplitude gives us a description of pion--nucleon scattering based on full  
QCD {\it and} QED. Lepton loops, e.g.\ vacuum polarization, have not been  
taken into account explicitly, but their effect is hidden in the electromagnetic  
counterterms. One could equally well compute lepton loops by using 
a Lagrangian containing photons and leptons as active degrees of freedom~\cite{knecht},  
but for our purposes the parameterization of these effects by counterterms  
is sufficient. Let us add a last remark about the relativistic form of our amplitude. 
The pions and photons are treated fully relativistically; on the other hand,  
since we restrict ourselves to the energy region just above threshold, we can consider  
the nucleons to be very heavy and work in heavy--baryon chiral perturbation  
theory~\cite{JM,BKKM}. This is a particular non--relativistic expansion of the amplitude,  
which amounts to organizing observables in a series of terms with increasing  
power of the inverse (heavy--) nucleon mass. 
 
\medskip\noindent 
Our calculations are based on the Lagrangian constructed
in~\cite{svenphd} (see also \cite{mms}). 
The relevant terms can be found in appendix~\ref{app:lagr}. 
 
\section{Fitting procedure 
\label{sec:fit}} 
 
After having constructed the full amplitude, we can proceed to fixing the  
unknown parameters. For this we use cross section data for elastic 
$\pi^+ p$ and $\pi^- p$ scattering as well as for the charge exchange  
$\pi^- p \to \pi^0 n$ reaction. Guided by the results of~\cite{FMS}, 
we decide to use experimental information up to pion laboratory momenta of $100$~MeV 
(i.e.\ $T_\pi \leq 32.2$~MeV) as input. We perform a combined fit 
of our isospin breaking amplitude to all three channels which have 
been studied experimentally. In the elastic channels, we  
use the data by Frank et al.~\cite{frank}, Brack et al.~\cite{brack}, and by Joram et al.~\cite{joram}. 
Low--energetic charge exchange data have been taken by 
Isenhower et al.~\cite{isenhower}, Salomon et al.~\cite{salomon}, and by Frlez et al.~\cite{frlez}. 
Note that we do not include the old data by Bertin et al.~\cite{bertin} and  
by Duclos et al.~\cite{duclos} in our data base, since these are nowadays 
generally believed to be erroneous. 
 
\medskip\noindent 
In the full amplitude, there appear 14 combinations of parameters 
which have to be fixed. Such a large number of unknowns 
makes a determination from pion--nucleon scattering data 
alone (almost) impossible. But, fortunately, some of these counterterms appear in 
other processes, which have been studied in great detail. 
The counterterms $c_{6/7}$, $d_{6/7}$, related to the 
coupling of a photon to the nucleon, can be fitted to the magnetic moments 
and the charge radii of proton and neutron (see e.g.~\cite{kubisff}). 
The separation of the nucleon mass difference into a strong and an 
electromagnetic piece~\cite{GL82,ms} fixes $c_5$ and $f_2$. 
We are now left with 11 parameters which have to be determined. 
 
\medskip\noindent 
We define the function which has to be minimized 
in analogy to the one introduced in~\cite{matsinew}, i.e.\  
\begin{equation} 
\chi^2  =  \sum_{j} w_j \chi_j^2~, 
\end{equation} 
with  
\begin{equation} 
\chi_j^2  =  \sum_{i=1}^{n_j} \frac{(y_{ij}^{\exp}-y_{ij}^{\rm th})^2}{(\sigma_{ij}^{\rm stat})^2}~, 
\end{equation} 
the partial contribution of the $j$th data set comprising $n_j$ entries, and 
the weights $w_j$ being given by 
\begin{equation} 
w_j = N (\sigma_j^{\rm sys})^{-2} / \sum_j  (\sigma_j^{\rm sys})^{-2}~. 
\end{equation} 
Here $N$ is the number of data sets we fit to, and the uncertainties 
are split into a statistical piece $\sigma_{ij}^{\rm stat}$ and 
a systematic contribution $\sigma_{j}^{\rm sys}$~\footnote{$\sigma_{j}^{\rm sys}$ is 
determined by the normalization uncertainty of the specific experiment.  
In analogy to~\cite{matsinew}, we attribute a systematic uncertainty  
of $5 \%$ to data sets for which this quantity is not given explicitly.}.  
The experimental data are denoted by $y_{ij}^{\exp}$, and we choose the LECs in such a way that 
our predicted cross sections $y_{ij}^{\rm th}$ minimize the $\chi^2$--function. 
 
\medskip\noindent 
We use the standard MINUIT minimization routines of the CERN library. 
Table~\ref{table:full_lecs} gives our best--fit values at the scale $\lambda = M_{\pi^0}$.  
Note that the quoted errors correspond to the MINUIT errors only and are certainly underestimated. 
The hadronic low--energy constants are of natural size, i.e.\ of order 1; as for 
the electromagnetic counterterms, some come out unnaturally large, but in this case 
they are accompanied by big uncertainties.  
It is obvious that one has to be careful about correlated 
parameters in a fit involving such a large number of unknowns; we indeed have  
large correlations among the strong--interaction counterterms. 
This does not come as a surprise since we restrict our analysis to a small 
energy region above threshold. There the pion energy, for example, is hardly distinguishable  
from the pion mass which causes the parameters accompanying these two structures to 
be strongly correlated. For these reasons we refrain from making predictions 
for  quantities like the sigma term or the pion--nucleon 
coupling constant $g_{\pi N N}$ which are determined by LEC combinations 
strongly correlated to other parameters. We have checked, however, that there are very small  
correlations between electromagnetic and strong counterterms\footnote{$\tilde{c}_1$  
needs special treatment in this respect. We will come back to this point in section~\ref{sec:strong}.}. 
This guarantees that we can reliably extract the strong phases from the full amplitude.  
The smallness of the $\chi^2$/dof value shows that we 
fit to data in a region where the third order amplitude is still precise enough. 
More important, it also demonstrates the consistency of the low--energy data base.

\renewcommand{\arraystretch}{1.2} 
\begin{table}[hbt] 
\begin{center} 
\begin{tabular}{|c|c|} 
    \hline 
    LEC      &  \\ 
\hline\hline 
$\tilde{c}_1= c_1 + e^2 F_\pi^2 f_1/(2 M_{\pi^0}^2)$                                & $( 0.64\pm 0.04)$~GeV$^{-1}$ \\ 
$c_2$                                                                               & $( 5.54\pm 0.05)$~GeV$^{-1}$ \\ 
$c_3$                                                                               & $(-5.34\pm 0.04)$~GeV$^{-1}$ \\ 
$c_4$                                                                               & $( 2.84\pm 0.15)$~GeV$^{-1}$ \\ 
\hline 
$\bar{d}_1+\bar{d}_2$                                                               & $(-2.24\pm 0.16)$~GeV$^{-2}$ \\ 
$\bar{d}_3$                                                                         & $( 0.81\pm 0.16)$~GeV$^{-2}$ \\ 
$\bar{d}_5$                                                                         & $( 0.67\pm 0.11)$~GeV$^{-2}$ \\ 
$\bar{d}_{14}-\bar{d}_{15}$                                                         & $(-0.63\pm 0.75)$~GeV$^{-2}$ \\ 
$\bar{d}_{18}$                                                                      & $(-10.14\pm 0.45)$~GeV$^{-2}$ \\ 
\hline 
$\tilde{g}_6 = \bar{g}_6 + \bar{g}_8 - \bar{k}_8/(128 \pi^2 F_\pi^2)-\bar{\ell}_6/(96 \pi^2 F_\pi^2)$   
                                                                                    & $(49.48\pm 10.15)$~GeV$^{-2}$\\  
$\tilde{g}_7 = \bar{g}_7 + \bar{k}_8/(64 \pi^2 F_\pi^2)$                            & $(55.07\pm 39.97)$~GeV$^{-2}$\\  
\hline 
$f_2$                        & $ -0.97$~GeV$^{-1}$*\\   
$B_0 (m_u - m_d) c_5$        & $ 0.00051$~GeV$^{-1}$*\\ 
$c_6$                        & $5.64$*\\  
$c_7$                        & $-2.88$*\\ 
$\bar{d}_6$                  & $0.38$~GeV$^{-2}$*\\ 
$\bar{d}_7$                  & $-0.65$~GeV$^{-2}$*\\ 
\hline\hline 
$\chi^2$/dof                 & $1.50 $ \\ 
\hline\hline 
\end{tabular} 
\caption{Values of the LECs resulting from a fit of the full 
         pion--nucleon amplitude to low--energy cross section data. 
         Parameter values marked by a * are input quantities. 
         \label{table:full_lecs}} 
\end{center} 
\end{table}

\section{Extraction of the strong amplitude 
\label{sec:strong}} 
 
After having determined all low--energy constants appearing  
in the pion--nucleon amplitude, we can now proceed to extracting  
the strong interaction piece. As pointed out in the introduction,  
it is this piece which is of particular interest for the determination  
of isospin breaking. In the previous paragraph we have used the full amplitude 
in order to describe pion--nucleon cross sections. The strong amplitude 
is defined as the QCD scattering amplitude with $m_u \neq m_d$, but $e^2 = 0$. 
We will thus not work in the isospin symmetry limit, but we will leave room for 
{\em strong} isospin breaking.  
 
\medskip\noindent 
In order to determine the strong amplitude, we have to know what the hadronic  
masses of the scattering particles are. We can write the charged and neutral  
pion masses as 
\bea 
M_{\pi^+}^2 = M_0^2\!\!\!\!&+  \frac{2}{F_\pi^2} e^2 C\!\!\!\!& + {\cal A}_+^{(4)} e^4 + {\cal B}_+^{(4)} e^2 (m_u - m_d) 
   + {\cal C}_+^{(4)} (m_u - m_d)^2~,\\ 
M_{\pi^0}^2 = M_0^2\!\!\!\!&                          &+ {\cal A}_0^{(4)} e^4 + {\cal B}_0^{(4)} e^2 (m_u - m_d) 
   + {\cal C}_0^{(4)} (m_u - m_d)^2~. 
\eea 
Here $M_0$ is the bare pion mass (more precisely, the leading term in the quark mass 
expansion of the pion mass in the two--flavor case with $m_s$ fixed),
and the ${\cal A}_{+/0}^{(4)}$, ${\cal B}_{+/0}^{(4)}$ and  
${\cal C}_{+/0}^{(4)}$ are combinations of low--energy constants from the fourth order  
mesonic Lagrangian with virtual photons. Note that the pion mass difference to leading order  
is of purely electromagnetic origin. We obtain the hadronic pion masses 
by switching off the electromagnetic interaction  
and using 
\bea 
|M_{\pi^0}^h - M_{\pi^0}| & \le & 1~{\rm MeV}~, \label{pi0had}\\ 
M_{\pi^+}^h & = & M_{\pi^0}^h +0.1~{\rm MeV}~.\label{piphad} 
\eea 
These numbers are based on the work by Knecht and Urech~\cite{KU98} using naturalness  
of the low--energy constants. 
 
\medskip\noindent 
As for the nucleon masses, we have 
\begin{equation} 
m_{p/n} = m - 4 M_{\pi^0}^2 c_1 \mp 2 B_0 (m_u - m_d) c_5 
- e^2 F_\pi^2 \left(f_1 + \frac{1 \pm 1}{2} f_2  + f_3\right)~, 
\end{equation} 
in terms of the bare nucleon mass $m$ (i.e. the nucleon mass in the
SU(2) chiral limit) and the counterterms given in appendix~\ref{app:lagr}. 
The hadronic masses will then be given by 
\begin{eqnarray} 
m_p^h & = & m - 4 M_{\pi^0}^2 c_1 - 2 B_0 (m_u - m_d) c_5 \no\\ 
      & = & m_p + e^2 F_\pi^2 (f_1 + f_2 + f_3)~, 
\label{prothad}\\ 
m_n^h & = & m - 4 M_{\pi^0}^2 c_1 + 2 B_0 (m_u - m_d) c_5 \no\\ 
      & = & m_n + e^2 F_\pi^2 (f_1 + f_3)~. 
\label{neuthad} 
\end{eqnarray} 
In the following we will take these mass differences into account, as well  
for the external particles as for the nucleons and pions in the intermediate  
states\footnote{This has not been done by the previous phase shift analyses  
which only work in the isospin symmetric limit. This is a severe  
drawback of these analyses, since the inclusion of such effects leads to 
important changes, as will be shown in the following.}. 
 
\medskip\noindent 
It becomes clear from eqs.~(\ref{prothad}) and (\ref{neuthad}) that we will not 
be able to determine the hadronic nucleon masses precisely, since we  
do not know the value of the LEC combination $(f_1 + f_3)$.  
An additional problem is due to the fact that, after including electromagnetic  
counterterms into the amplitude, we can no more determine $c_1$ and $f_1$  
separately, as only the combination  
$\tilde{c}_1 = c_1 + e^2 F_\pi^2 f_1/(2 M_{\pi^0}^2)$ appears. 
Since we do not know the precise value of 
these LECs, all we can do is to assume natural sizes for $f_1$ and $f_3$; 
we will thus vary these parameters over the interval 
$f_{1/3} = (0 \pm 1)$~GeV$^{-1}$~\cite{fmi}.  
Our results will, however, turn out to be insensitive to the  
precise values of these parameters. Also the lack of knowledge of the absolute 
value of the hadronic pion masses (see eq.~(\ref{pi0had})) does not affect our results;  
only the mass differences are relevant.

\begin{figure}[p] 
\centerline{ 
\epsfysize=7in 
\epsffile{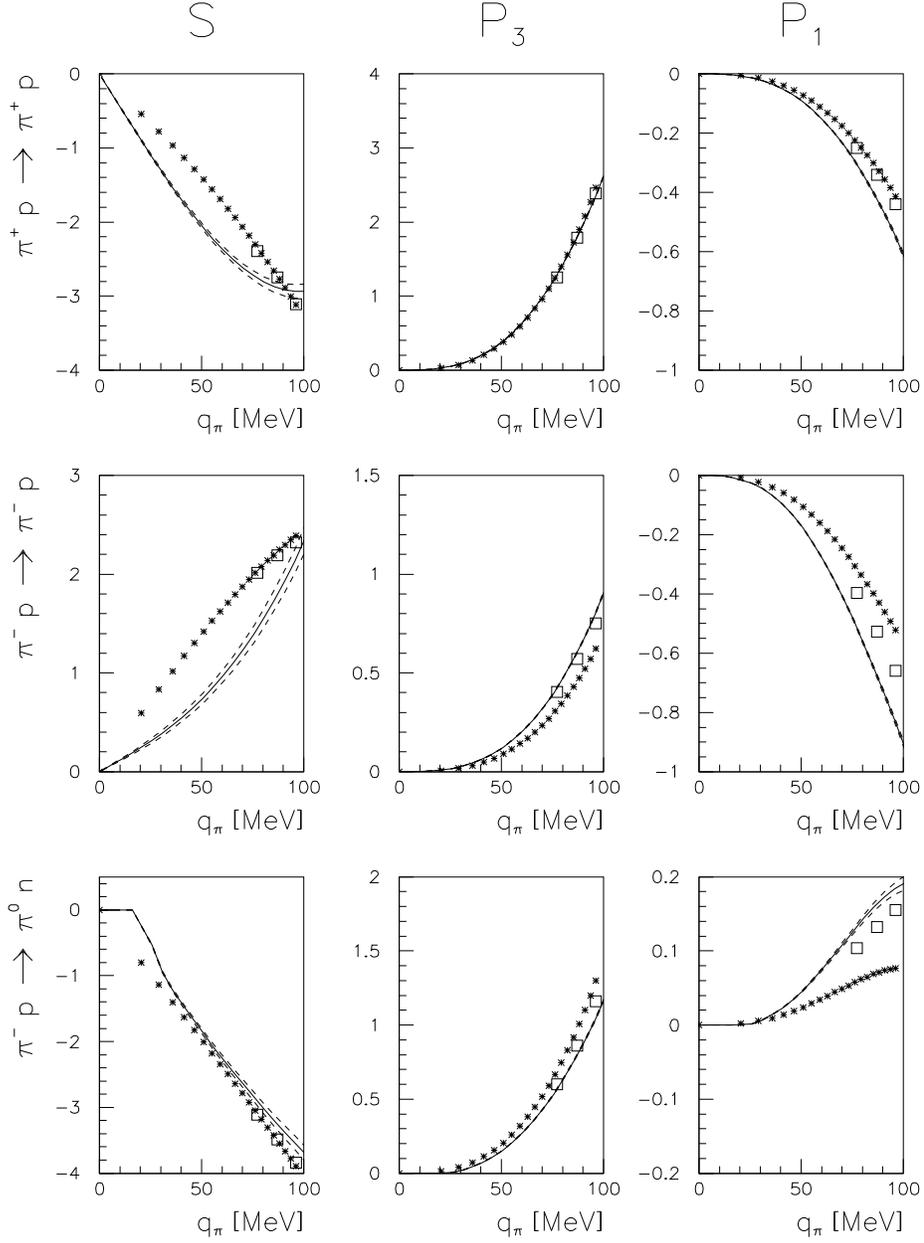} 
} 
\vskip 0.5cm 
\caption{ 
Strong pion--nucleon phase shifts as a function of the 
pion laboratory momentum $q_\pi$ for the three measured channels. 
The solid line corresponds to our solution, the dashed one 
to the one--sigma uncertainty range. Also shown are the 
EM98~\cite{mats} (stars) and the EM00~\cite{matsinew} (open squares) phases. 
} 
\label{fig:phasesem} 
\end{figure} 
 
\begin{figure}[p] 
\centerline{ 
\epsfysize=7in 
\epsffile{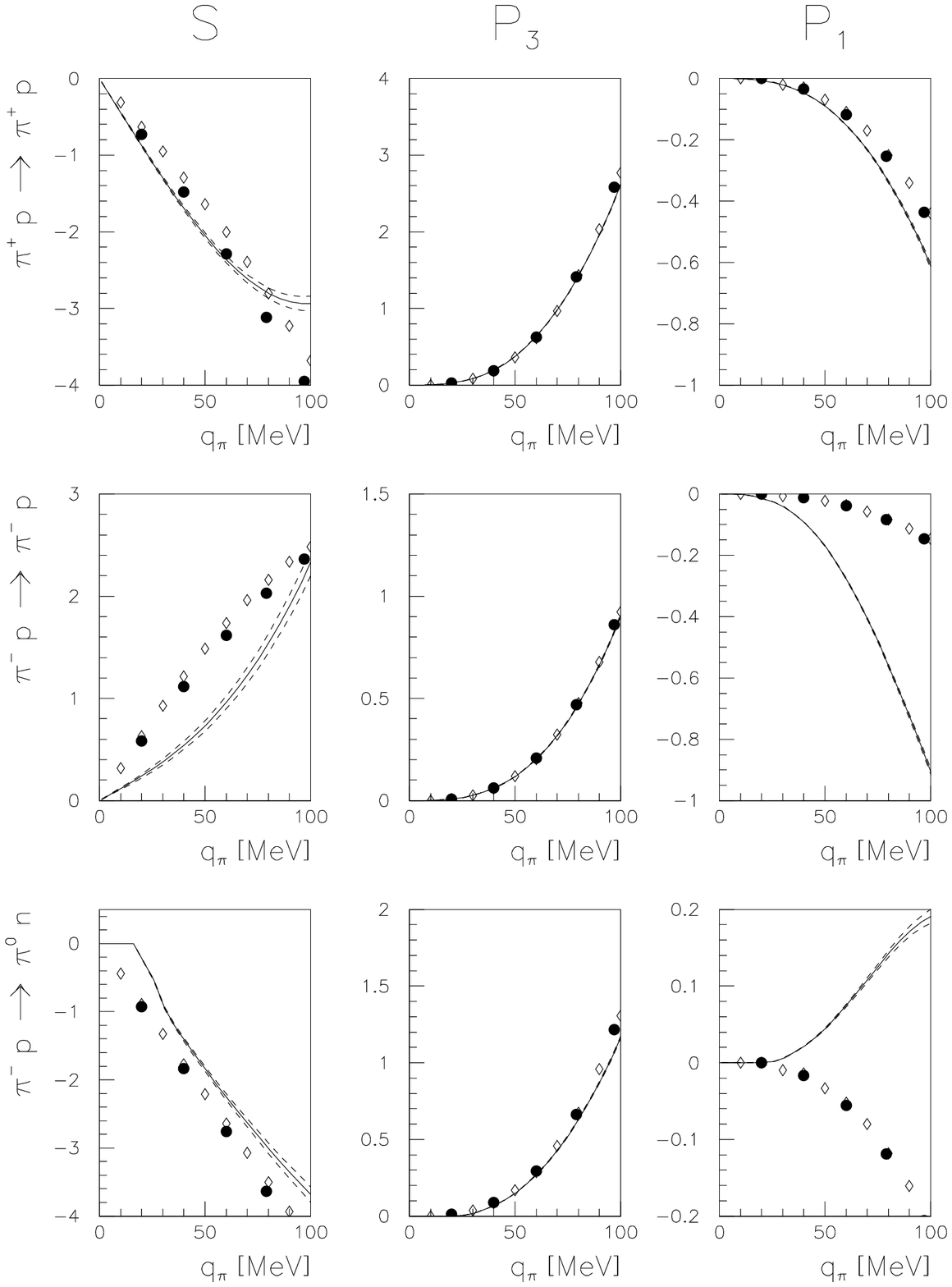} 
} 
\vskip 0.5cm 
\caption{ 
Strong pion--nucleon phase shifts as a function of the 
pion laboratory momentum $q_\pi$ for the three measured channels. 
The solid line corresponds to our solution, the dashed one 
to the one--sigma uncertainty range.  Also shown are the 
KA85~\cite{ka85} (full dots) and the SP98~\cite{sp98} (open diamonds) phases.   
} 
\label{fig:phaseskasp} 
\end{figure} 
 
\medskip\noindent 
Fig.~\ref{fig:phasesem} shows our prediction for the strong phases as compared 
to the phase shift analyses performed by Matsinos et al.~\cite{mats,matsinew} which do not contain 
isospin violating contributions. The range of uncertainty in our prediction, 
resulting from the correlated error bars of the low--energy constants, 
is given by the dashed curves. In the P$_3$--waves, the agreement between both analyses is perfect. 
In the P$_1$--waves, our solution is closer to the recent analysis EM00. 
The most dramatic discrepancy lies in the S--waves of the elastic channels;  
this difference is not due to the inclusion of the isospin violating effects, 
but is already present in the isospin symmetric part of the amplitude.  
The kink in the S--wave phase shift for the charge exchange channel  
comes from the fact that the ingoing pion needs to have a lab momentum of  
$\sim 25$~MeV in order to produce the heavier final state. 
(Remember that only the hadronic mass difference of the involved particles is of relevance.) 
In fig.~\ref{fig:phaseskasp} we show our solution for the phase shifts compared to the  
KA85~\cite{ka85} and the SP98~\cite{sp98} analyses. The agreement in the P$_3$--waves  
is still very good, whereas there are some discrepancies in the P$_1$ phases. 
Our $\pi^+ p$ S--wave phases agree better with the KA85 and SP98 ones than with the EM98(00) 
solution. But there remains a large difference in the S--wave $\pi^- p$ phase shift. 
 
\begin{figure}[htb] 
\vskip 0.5cm 
\centerline{ 
\epsfysize=1.8in 
\epsffile{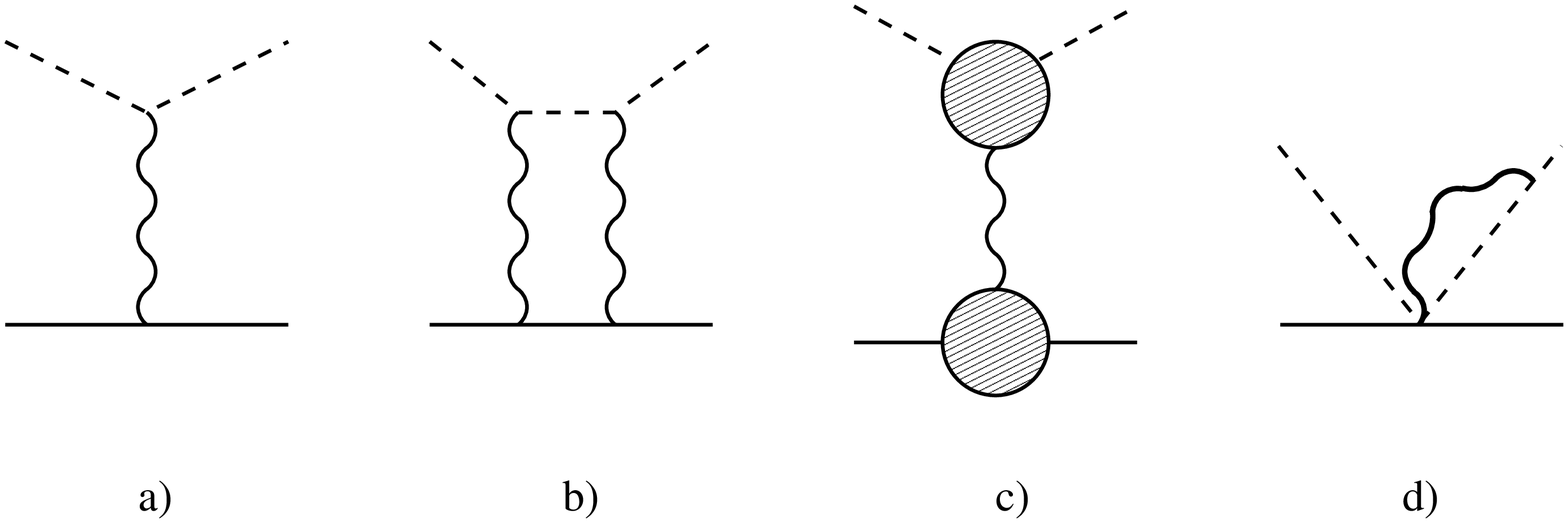} 
} 
\vskip 0.5cm 
\caption{ 
a)~One--photon exchange diagram, b)~two--photon exchange diagram,  
c)~vertex corrections to the one--photon exchange, 
d)~typical graph involving non--linear pion--nucleon--photon couplings. 
The solid, dashed, and wiggly lines denote nucleons, pions, and photons, in order} 
\label{fig:diagamma} 
\end{figure}

\begin{figure}[p] 
\centerline{ 
\epsfysize=7in 
\epsffile{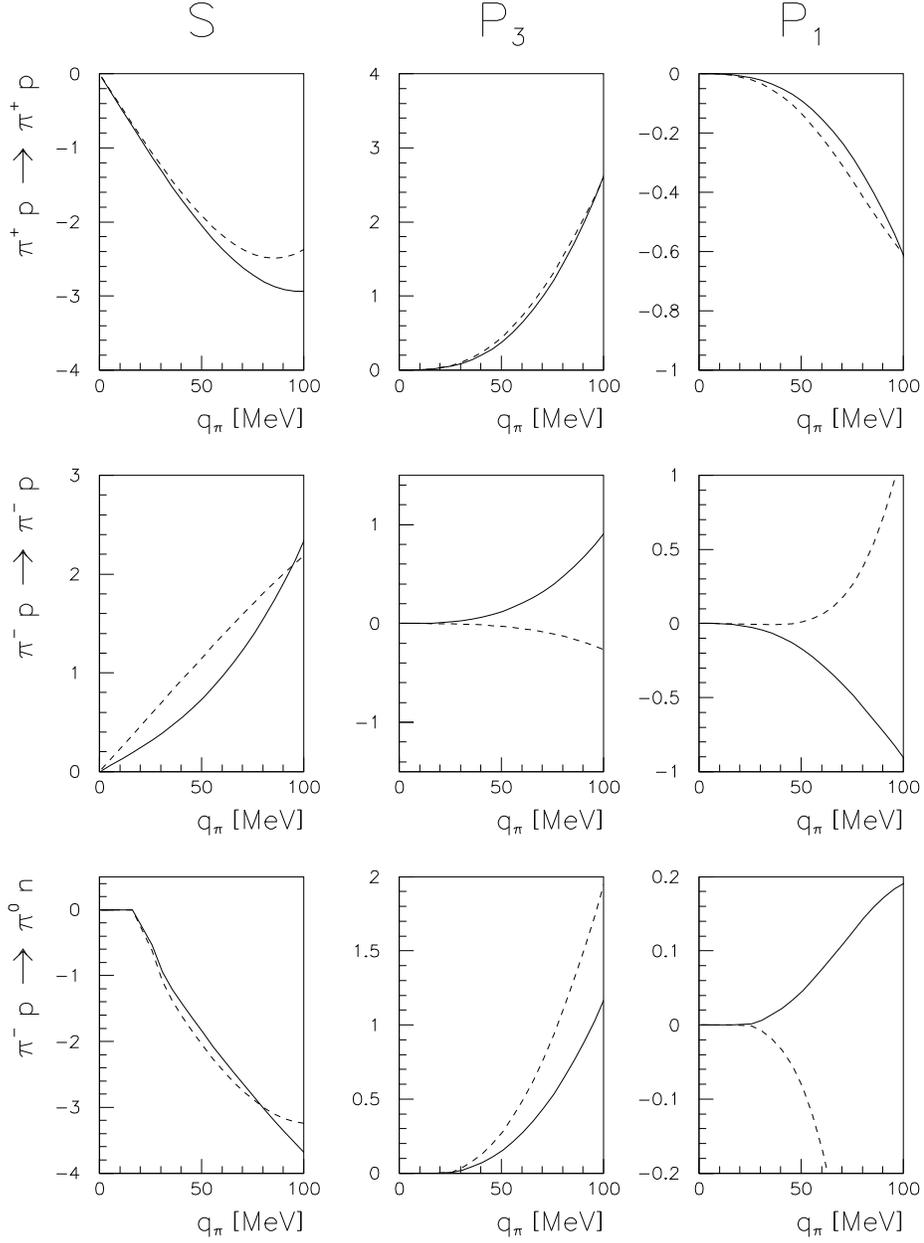} 
} 
\vskip 0.5cm 
\caption{ 
Strong pion--nucleon phase shifts as a function of the pion laboratory 
momentum $q_\pi$. The full line is deduced from an analysis based on the 
complete pion--nucleon amplitude. To obtain the dashed line, we have fitted  
the strong amplitude together with the one--photon exchange diagram with dressed 
vertices and the two--photon exchange process to the data, and then switched off  
the electromagnetic interaction (see text). 
} 
\label{fig:phases_gamma} 
\end{figure}

\medskip\noindent 
In order to explain where the difference in the S--waves comes from, we further analyze  
the contributions to the electromagnetic corrections (see fig.~\ref{fig:phases_gamma}).  
The strong phase shifts shown in figs.~\ref{fig:phasesem}, \ref{fig:phaseskasp} 
are obtained when fitting the full pion--nucleon amplitude with virtual photons to  
the experimental data, and then switching off the electromagnetic coupling. This is  
also shown by the solid line in fig.~\ref{fig:phases_gamma}. 
The dashed line, on the other hand, is obtained if we do not include all 
virtual--photon diagrams, but only the one--photon exchange diagram with dressed 
vertices and the two--photon exchange process (see diagrams a)--c) in fig.~\ref{fig:diagamma}). 
The way to treat the one--photon and two--photon exchange between the charged  
pion and the proton is standard. The inclusion of corrections to the $N \gamma$  
and the $\pi \gamma$ vertex does not lead to big changes in the phases.  
The remaining virtual--loop diagrams (one example of which is drawn in d) in  
fig.~\ref{fig:diagamma}) have not been included 
when drawing the dashed line. We think that such diagrams can only be accounted for correctly 
when the strong interaction and electromagnetism are treated in a consistent way.  
These effects are very large and cause this unusual bending of the $\pi^- p$ S--wave phase shift. 
Previous phase shift analyses have obviously underestimated these contributions. 
 
\section{Isospin violation of the strong interactions 
\label{sec:isoviol}} 
 
The aim of this section is to determine the size of isospin violation in 
low--energy pion--nucleon scattering. Isospin is an approximate symmetry of the 
strong interactions, for electromagnetism it is largely broken. Consequently, it 
does not make sense to study isospin violation of the full pion--nucleon amplitude. 
The interesting part lies in the strong amplitude, which we have isolated  
in section~\ref{sec:strong}. As in section~\ref{sec:strong}, we will switch off every 
electromagnetic interaction; we will thus be left with the isospin violating 
two-pion--nucleon vertex $\sim c_5$, and the hadronic mass differences of the nucleons 
and the pions. 
 
\medskip\noindent 
In order to quantify isospin breaking, we will employ the usual triangle ratio~\cite{kg,gibbs,matsi,fmi} 
\begin{equation} 
R = 2 \, \frac{f_{\pi^+ p \to \pi^+ p} - f_{\pi^- p \to \pi^- p} - \sqrt{2} \, f_{\pi^- p \to \pi^0 n}}  
          {f_{\pi^+ p \to \pi^+ p} - f_{\pi^- p \to \pi^- p} + \sqrt{2} \, f_{\pi^- p \to \pi^0 n}}~. 
\end{equation} 
Here $f$ stands for the strong part of the pion--nucleon amplitude. In the following, we will always  
form the ratio $R$ with the real part of the scattering amplitude\footnote{In chiral perturbation  
theory, unitarity is only perturbatively fulfilled. Since imaginary parts are due to 
loop diagrams which only come in at third order, they are known with less precision than 
the real parts of the amplitudes.}. In case that isospin is a good symmetry $R$ vanishes 
over the whole energy range.  
 
\medskip\noindent 
It has been argued in~\cite{fmi} that it is advantageous not to consider the usual 
P$_3$-- and P$_1$--wave projections, but to focus instead on the P--wave parts proportional to 
the spin--flip and the spin--non--flip amplitudes $h$ and $g$. These are linked to the usual $T$--matrix 
elements via 
\bea\label{pin_defamp}  
T_{\pi N} &=& \sqrt{\frac{E_1+m_1}{2 m_1}} \sqrt{\frac{E_2+m_2}{2 m_2}}  
\Big[ g  + i \vec \sigma \cdot(\vec{q}_2\times \vec{q}_1\,) \, h \Big]~. 
\eea 
$E_{1/2}$ is the energy of the in--/outgoing nucleon with mass $m_{1/2}$, the in--/outgoing pions 
have four--momenta $q_{1/2}$. 
The usual S-- and P--waves are then given by 
\bea  
{\rm S} (s) &=&    {{\sqrt{E_1+m_1} \sqrt{E_2+m_2}} \over 16 \pi \sqrt{s}} {\cal S}(s)~, \\  
{\rm P}_3 (s) &=&   {{\sqrt{E_1+m_1} \sqrt{E_2+m_2}} \over 16 \pi \sqrt{s}} \, |\vec{q}_1| |\vec{q}_2| 
\Big( {\cal G}(s) + {\cal H}(s)  \Big)~, \\  
{\rm P}_1 (s) &=&  {{\sqrt{E_1+m_1} \sqrt{E_2+m_2}} \over 16 \pi \sqrt{s}}  \, |\vec{q}_1| |\vec{q}_2|  
\Big( {\cal G}(s) -2 {\cal H}(s) \Big)~,   
\eea  
with  
\bea  
{\cal S}(s) & = & \int_{-1}^{+1} dz \, g(s,z)~,\\  
{\cal G}(s) & = & \frac{1}{|\vec{q}_1| |\vec{q}_2|}\int_{-1}^{+1} dz \, z\, g(s,z)~,\\  
{\cal H}(s) & = &   \int_{-1}^{+1} dz \, \, \frac{z^2-1}{2} \, h(s,z)~.  
\eea  
The integration in the last equations is performed over the cosine of the scattering angle $z$. 
In the following we will consider isospin violation in the ${\cal S}$-- as well as in  the ${\cal G}$-- 
and ${\cal H}$--projections of $R$. 
 
\begin{figure}[hbt] 
\centerline{ 
\epsfysize=7in 
\epsffile{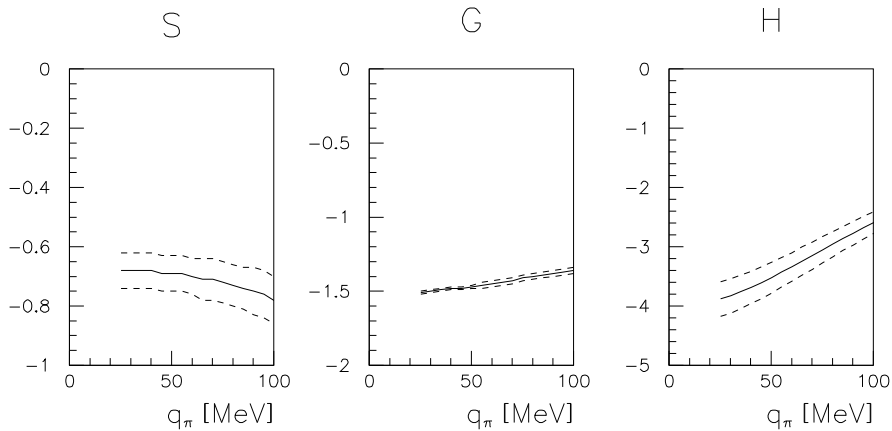} 
} 
\vskip -12cm 
\caption{ 
$R [\%]$ as a function of the pion laboratory momentum $q_\pi$. The dashed lines indicate  
the one--sigma range of our prediction.} 
\label{fig:triangle} 
\end{figure}

\medskip\noindent 
There is one additional problem: as pointed out before, the threshold of the charge  
exchange reaction lies at $q_\pi \sim 25$~MeV. Below this value, the SCX channel is closed 
and the triangle relation does not make sense. Nevertheless, the different threshold 
energies of the elastic and the inelastic channels are a consequence of  
strong isospin breaking (remember that only the hadronic mass difference of the  
particles has been included). Therefore, we want to take this effect into 
account; in fig.~\ref{fig:triangle} we show the momentum dependence of $R$ (in $\%$) 
for momenta $q_\pi$  above 25~MeV. In this energy domain all three channels are open.  
Again, the dashed line indicates the one--sigma uncertainty range of our prediction. 
For the ${\cal S}$--projection, isospin breaking lies around $-0.7 \%$.  
Also the ${\cal G}$--projection varies very little with energy, breaking in this channel 
is somewhat larger, $\sim -1.5 \%$. The biggest effect is found to be in 
${\cal H}$ where isospin violation lies between $-4$ and $-2.5 \%$ for $q_\pi $ below 100~MeV. 
 
\begin{figure}[hbt] 
\centerline{ 
\epsfysize=7in 
\epsffile{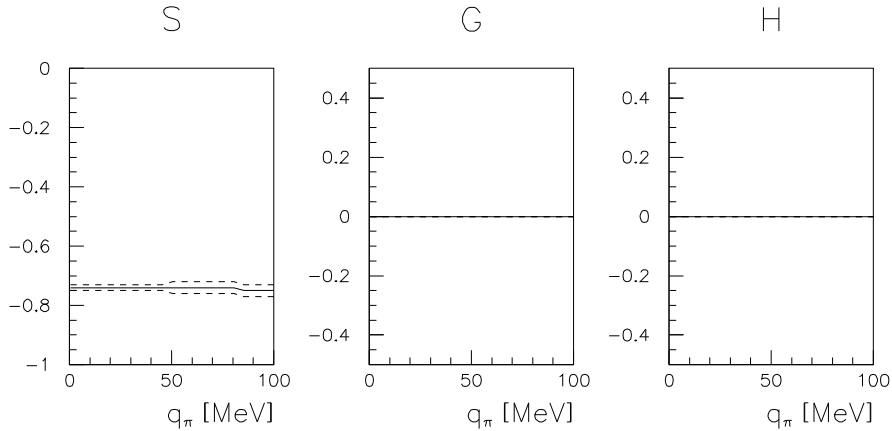} 
} 
\vskip -12cm 
\caption{ 
$R [\%]$ as a function of the pion laboratory momentum $q_\pi$ in the scenario where 
$m_p^h = m_n^h$ and $M_{\pi^0}^h = M_{\pi^+}^h$, as described in the text.  
The dashed lines indicate the one--sigma range of our prediction.} 
\label{fig:triangle_id} 
\end{figure}

\medskip\noindent 
Instead we could also argue that we do not want to take threshold effects into account, and that we 
want to separate ``static'' (i.e.\ related to kinematic effects) from ``dynamical''  
(i.e.\ due to isospin violating pion--nucleon couplings) isospin breaking~\cite{aron}.  
In fig.~\ref{fig:triangle_id}, we consider the case where the hadronic mass of the  
proton and the neutron, as well as the hadronic mass of the pions, are taken to be equal 
($m_p^h = m_n^h = (m_p + m_n)/2, M_{\pi^0}^h = M_{\pi^+}^h = M_{\pi^0}$).  
There is thus no static isospin breaking, but we only show effects due to the two-pion--nucleon vertex 
proportional to $c_5$. This coupling only shows up in the S--wave of the charge exchange reaction; 
it causes isospin to be broken by $-0.75 \%$ in ${\cal S}$. There is {\it no} dynamical isospin breaking in the 
P--waves (to the order considered here). 
 
\section{Summary and Outlook} 
For a reliable determination of the size of isospin breaking in the strong interactions, 
it is primordial to describe electromagnetic and strong effects consistently.  
In the present work this is achieved by an analysis of pion--nucleon scattering in  
the framework of chiral perturbation theory. $\chi$PT does not leave any doubt about the correct 
definition of the hadronic masses of pions and nucleons, and allows 
to extract the strong part of the scattering amplitude in a unique way. 
After determining the unknown low--energy constants by a fit to experimental data, we 
switch off all electromagnetic interactions and describe QCD with unequal  
up-- and down--quark masses and $e^2 = 0$. The so--determined strong phase shifts 
agree with those of Matsinos et al.~\cite{mats,matsinew} in the P--waves, but we find 
a sizeably different behavior in the S--waves. We trace this difference back to the  
inclusion or omission of non--linear photon--pion--nucleon couplings. 
 
\medskip\noindent 
We address the question of isospin violation by studying the usual triangle 
relation involving elastic $\pi^\pm p$ scattering and the charge exchange reaction. 
An important advantage of the $\chi$PT calculation lies in the fact that we can  
easily separate dynamical from static isospin breaking. Dynamical isospin breaking 
only occurs in the S--wave and is very small, $\sim 0.75 \%$. Static effects do not  
increase the size of isospin violation in the S--wave significantly; by no means can we account for 
the reported 6 -- 7 $\%$ isospin breaking~\cite{gibbs,matsi}. 
 
\medskip\noindent 
We have found large error bars on our parameter values. In order to improve 
this situation, we would like to fit to more experimental data. However,  
a third order $\chi$PT calculation allows to describe scattering data 
for pion laboratory momenta not much higher than 100~MeV, a region where the data situation  
is not yet as good as one would hope. A fourth order calculation would certainly allow 
to fit to data higher in energy, but, on the other hand, would also introduce many more 
unknown coupling constants. Since isospin breaking effects are expected to be most  
prominent in the energy region we consider in this work, we do not judge it very  
promising to extend the analysis to full one--loop (fourth) order. Additional data  
for pion--nucleon scattering at very low energies would be very helpful in this respect.  
Also a combined fit to several reactions involving nucleons, pions, and photons, 
e.g.\ pion electro-- and photoproduction, as well as $\pi N \to \pi \pi N$, would help in  
pinning down the fundamental low--energy constants more precisely.

\section*{Acknowledgments} 
 
We would like to thank Bastian Kubis for many helpful discussions and a  
careful reading of the manuscript. 
 

\appendix

\section{Lagrangian 
\label{app:lagr}} 
\def\theequation{\Alph{section}.\arabic{equation}} 
\setcounter{equation}{0} 
 
In this appendix we give the relevant counterterms, appearing in the pion--pion  
and pion--nucleon Lagrangians. We only display the finite parts. The complete  
set of terms can be found in~\cite{svenphd}.

\medskip\noindent 
Let us start with the Lagrangian involving pions, nucleons and virtual photons.  
We only give the form of the counterterms, the expression for the terms with 
fixed coefficients is canonical. At second and third order, these read 
\bea 
{\cal L}_{\gamma^* \pi N}^{(2)} &= \bar{N}  &\Bigg[ 
c_1 \lanl \chi_+ \ranl  
+ \frac{c_2}{2}  \lanl (v\cdot u)^2\ranl 
+\frac{c_3}{2} \lanl u^2 \ranl 
+\frac{c_4}{2}  [S^\mu,S^\nu] [u_\mu,u_\nu]\no\\&& 
+c_5 \hat{\chi}_+  
-\frac{i c_6}{4m} [S^\mu,S^\nu] F_{\mu\nu}^+ 
-\frac{i c_7}{4m}  [S^\mu,S^\nu] \lanl F_{\mu\nu}^+\ranl\no\\&& 
+f_1 F_0^2 \lanl Q_+^2 - Q_-^2 \ranl 
+ f_2 F_0^2 \lanl Q_+ \ranl Q_+ 
+ f_3 F_0^2 \lanl Q_+ \ranl ^2 
+ \ldots 
\Bigg] N~, \\ 
{\cal L}_{\gamma^* \pi N}^{(3)} &= \bar{N}  &\Bigg[ 
\frac{c_2}{2m} \big( i \lanl v\cdot u u_\mu \ranl D^\mu + {\rm h.c.} \big) 
+\frac{c_4}{4 m} \big( \epsilon^{\mu\nu\alpha\beta} v_\alpha S_\beta [u_\mu, u_\nu] v\cdot D + {\rm h.c.} \big)\no\\&& 
+\frac{c_4}{4m} \big( \epsilon^{\mu\nu\alpha\beta} S_\alpha [u_\mu, u_\nu] D_\beta + {\rm h.c.} \big) 
+\frac{i c_4}{4m} [v\cdot u,[D^\mu,u_\mu]]\no\\&& 
-\frac{c_6}{8 m^2} \big( i \epsilon^{\mu\nu\alpha\beta} v_\alpha S_\beta \hat{F}_{\mu\nu}^+ v\cdot D + {\rm h.c.} \big) 
-\frac{c_6}{8 m^2} \big( i \epsilon^{\mu\nu\alpha\beta} S_\alpha \hat{F}_{\mu\nu}^+ D_\beta + {\rm h.c.} \big)\no\\&& 
-\frac{c_6 + 2 c_7}{16 m^2} \big( i \epsilon^{\mu\nu\alpha\beta} v_\alpha S_\beta \lanl F_{\mu\nu}^+ \ranl v\cdot D + {\rm h.c.} \big) 
-\frac{c_6 + 2 c_7}{16 m^2} \big( i \epsilon^{\mu\nu\alpha\beta} S_\alpha \lanl F_{\mu\nu}^+ \ranl D_\beta + {\rm h.c.} \big)\no\\&& 
+i \bar{d}_1 [u_\mu,[v\cdot D,u^\mu]] 
+i \left( \bar{d}_2-\frac{c_4}{4 m} \right) [u_\mu,[D^\mu,v\cdot u]] 
+ i \bar{d}_3[v\cdot u,[v\cdot D,v\cdot u]]\no\\&& 
+\bar{d}_5 [\chi_-,v\cdot u] 
+\left( \bar{d}_6 -\frac{c_6}{8 m^2}\right) [D^\mu, \hat{F}_{\mu\nu}^+] v^\nu 
+\left( \bar{d}_7-\frac{c_6 + 2 c_7}{16 m^2}\right) [D^\mu, \lanl F_{\mu\nu}^+ \ranl] v^\nu\no\\&& 
+\bar{d}_{14} \epsilon^{\mu\nu\alpha\beta} v_\alpha S_\beta \lanl[v\cdot D,u_\mu]u_\nu\ranl 
+\bar{d}_{15} \epsilon^{\mu\nu\alpha\beta} v_\alpha S_\beta \lanl u_\mu [D_\nu, v\cdot u]\ranl\no\\&& 
+\bar{g}_6 F_0^2 S^\mu \lanl Q_- v\cdot u\ranl \lanl Q_+\ranl 
+\bar{g}_7 F_0^2 \lanl Q_+ v\cdot u\ranl Q_-\no\\&& 
+\bar{g}_8 F_0^2 \lanl Q_- v\cdot u\ranl Q_+ 
+ \ldots 
\Bigg] N~. 
\eea 
This heavy--baryon form of the Lagrangian is given in terms of the nucleon velocity $v_\mu$ and the 
spin operator $S_\mu = \frac{i}{2} \gamma_5 \sigma_{\mu\nu} v^\nu$. $\epsilon$ is the totally antisymmetric  
tensor in four dimensions, and we choose the convention $\epsilon^{0123}=-1$.  
$F_0$ is the unrenormalized pion decay constant, $N$ is the nucleon spinor, 
and ${\rm h.c.}$ stands for the hermitian conjugate of a particular term. 
Moreover, $\lanl A \ranl$ denotes the flavor trace of $A$, and $\hat{A}=A-\lanl A \ranl/2$ its traceless part. 
 
\medskip\noindent 
Let us restrict ourselves to the case in which there are no external sources but where virtual photons are included. 
These are parameterized by the electromagnetic field $A_\mu$ and the field strength  
tensor $F_{\mu\nu} = \partial_\mu A_\nu - \partial_\nu A_\mu$.  
The elementary building blocks of the Lagrangian are given in terms of 
the field 
\bea
u&=& \sqrt{U} =
\left(\sqrt{1-\frac{\vec{\pi}^2}{F^2}} + i \frac{\vec{\tau}\cdot \vec{\pi}}{F}\right)^{1/2}~, 
\eea
and the nucleon charge matrix $Q$ by
\bea 
u_{\mu} & = & i\{u^{\dagger}(\partial_{\mu}+i Q A_\mu)u-u(\partial_{\mu}+i Q A_\mu)u^{\dagger}\} ~,\no\\ 
\chi_{\pm} & = & u^{\dagger}\chi u^{\dagger}\pm u\chi^{\dagger}u~, \nonumber\\ 
Q_\pm & = & \frac{1}{2} \{u^\dagger Q u \pm u Q u^\dagger\}~,\no\\ 
F^{\pm}_{\mu\nu} & = & -2 Q_\pm F_{\mu\nu}~. 
\eea 
$\chi$ is related to the quark mass matrix, $\chi = 2 \, B_0 \, {\rm diag}(m_u, m_d)$, 
where $B_0 = - \lanl 0 | \bar{q} q | 0 \ranl / F_0^2 + \ldots $. We assume 
$B_0 \gg F_0$, i.e.\ we work in standard $\chi$PT. 
The covariant derivative reads  
\bea 
[D_\mu, A] &=& \partial_\mu A + [\Gamma_\mu, A]~,\no\\ 
\Gamma_\mu & = & \frac{1}{2} \{ u^\dagger (\partial_\mu + i Q A_\mu) u + u (\partial_\mu + i Q A_\mu) u^\dagger \}~. 
\eea 
 
\medskip\noindent 
In the pion--photon sector, we only need few terms which enter the amplitude via renormalization.  
In the Feynman gauge, we obtain 
\bea 
{\cal L}_{\gamma^* \pi \pi}^{(2)} &=& 
-\frac{1}{4} F_{\mu\nu} F^{\mu\nu}  
+\frac{F_0^2}{4} \lanl u_\mu u^\mu + \chi_+ \ranl 
+ C F_0^2 \lanl Q_+^2 - Q_-^2 \ranl~,\\ 
{\cal L}_{\gamma^* \pi \pi}^{(4)} &=& 
-\frac{\bar{\ell}_6}{384 \pi^2} \lanl F_{\mu\nu}^+ [u^\mu,u^\nu] \ranl 
+\frac{\bar{k}_8}{64 \pi^2} \lanl u^\mu [Q_-,i c_\mu^+]+u^\mu[Q_+,i c_\mu^-] \ranl 
+ \ldots 
\eea 
Here, $c_\mu^\pm = -\frac{i}{2} \{ u^\dagger [r_\mu, Q] u \pm u [l_\mu, Q] u^\dagger \}$ is needed  
for the renormalization of $F_\pi$.  
 

\end{document}